\documentclass[preprint,12pt]{elsarticle}

\usepackage{amssymb}

\journal{Physics Letters A}
\def\mib#1{\mbox{\boldmath$#1$\unboldmath}}
\def\mibsub#1{\mbox{\footnotesize \boldmath$#1$\unboldmath}}
\def\xi{\mib{x}_{\mbox{\scriptsize i}}}
\def\ti{t_{\mbox{\scriptsize i}}}
\def\xf{\mib{x}_{\mbox{\scriptsize f}}}
\def\tf{t_{\mbox{\scriptsize f}}}
\def\psif{\psi_{\mbox{\scriptsize f}}}
\def\psii{\psi_{\mbox{\scriptsize i}}}
\def\sx{\sigma_x}
\def\sz{\sigma_z}
\def\s0{\sigma_0}

\def\la{\langle}
\def\ra{\rangle}
\def\e{\mbox{e}}
\def\ket#1{|#1\ra}
\def\bra#1{\la#1|}
\def\ua{\uparrow}
\def\da{\downarrow}

\def\DF#1#2{\frac{\partial #1}{\partial #2}}
\def\DDF#1#2{\frac{\partial^2 #1}{{\partial#2}^2}}
\begin{document}

\begin{frontmatter}

\title{Weak value analogue in classical stochastic process}

\author{Hiroyuki Tomita}
\ead{tomita@alice.math.kindai.ac.jp}
\address{Research Center for Quantum Computing,\\
Kinki University,\\
Kowakae 3-4-1, Higashi-Osaka, 577-8502, Japan}
\begin{abstract}
The time evolution of the two-time conditional probability
of the classical stochastic process is described in an analogous
form of the quantum mechanical wave equations.
By using it, we emulate the same strange behaviors
as those of the weak value in the quantum mechanics.
A negative probability and abnormal expectations of some quantities
remarkablely larger than their inherent norms are found
in an example of a stochastic Ising spin system.
\end{abstract}
\begin{keyword}
Weak value \sep Stochastic process \sep Two-time conditional probability
\sep Stochastic Ising model


\end{keyword}

\end{frontmatter}

\section{Introduction}
A notion of the weak value proposed by Aharonov et al \cite{Aharonov}
 has brought a new 
understanding on the quantum observation, i.e. a weak measurement 
\cite{Measurement} which hardly disturbs the quantum state.
The reason of this strange nature of the quantum measurement is that
the weak value is defined as an expectation with the condition
of two-time observations of the initial and the final states which 
differ from one another.
This condition is very rare case with little probability
and is far from the main behavior of a given quantum system.
Then the observation of the weak value does not disturb the quantum system
not so fatally.
As a result of this rather fictitious probability, 
the weak value happens to be abnormally enhanced from its
inherent norm.
\par
The purpose of this letter is to make the mechanism of this
abnormal behavior clearer by using a classical stochastic model,
in which we can avoid the ambiguity of the complex probability
in the quantum case \cite{A-B}.
\par
We introduce a conventional transformation of the stochastic master equation
to a self-adjoint form in the following section.
A good analogy with the quantum mechanics is found by applying this
tarnsformation to the two-time conditional probability (TTCP). 
This is shown in Sec.3.
An example of the stochastic Ising model which shows an abnormal 
enhancement of the expectations of some quantities with respect to TTCP
is given in Sec.4.
In Sec.5 we discuss an extention of TTCP to a density matrix form
to complete the analogy with the quantum mechanics.
The last section is devoted to brief summary and discussions.

\section{Self-adjoint form of stochastic master equation}
First let us survey the well-known transformation \cite{KMK} 
to a self-adjoint form of the stochastic master equation.
\par
Let \mib{x} be a set of stochastic variables described by a time-dependent
conditional probability, 
$P(\mib{x},t|\xi,\ti)$ for $t\ge \ti$, which obeys the following 
stationary, Markoffian master equation, i.e. the Chapman-Kolmogorov
\textit{forward} equation,
\begin{eqnarray} \label{Master}
\frac{\partial}{\partial t}P(\mib{x},t|\xi,\ti)
&=&-\sum_{\mibsub{x'}}W(\mib{x}\to\mib{x'})P(\mib{x},t|\xi,\ti)
+\sum_{\mibsub{x'}}W(\mib{x'}\to\mib{x})P(\mib{x'},t|\xi,\ti)
\nonumber\\
&=&-\sum_{\mibsub{x'}}L(\mib{x},\mib{x'})P(\mib{x'},t|\xi,\ti),
\end{eqnarray}
where
\[
L(\mib{x},\mib{x'})=\left[\sum_{\mibsub{x''}}
W(\mib{x}\to\mib{x''})\right]\delta(\mibsub{x}-\mibsub{x'})
-W(\mib{x'}\to\mib{x}).
\]
The matrix $L$ has an eigenvalue $\lambda_0=0$ 
correspondding to the steady state,
\[
P_0(\mib{x})=\lim_{t-\ti\to\infty} P(\mib{x},t|\xi,\ti).
\]
Let us introduce a \textit{wave function} related to this
forward conditional probability by
\begin{equation}
\psi(\mib{x},t|\xi,\ti)=\phi_0(\mib{x})^{-1}
P(\mib{x},t|\xi,\ti),~~(t\ge \ti)
\end{equation}
where $\phi_0(\mib{x})=P_0(\mib{x})^{1/2}$.
This function $\psi$ obeys the forward wave equation,
\begin{equation} \label{Forward}
\DF{}{t}\psi(\mib{x},t)=-\sum_{\mibsub{x'}}
H(\mib{x},\mib{x'})\psi(\mib{x'},t),
\end{equation}
where $H$ is defined by
\begin{equation}
H(\mib{x},\mib{x'})=\phi_0(\mib{x})^{-1}L(\mib{x},\mib{x'})
\phi_0(\mib{x'}).
\end{equation}
For the time being the initial condition $(\xi,\ti)$ in $\psi$ is
abbreviated.
The function $\phi_0(\mib{x})$ is an eigenfunction of Eq.(\ref{Forward})
for $\lambda_0=0$.
\par
The merit of this transformation is that the eigenvalue problem of a given
master equation is simplified, if the matrix $H$ is symmetric, i.e.
\[
H(\mib{x},\mib{x'})=H(\mib{x'},\mib{x}).
\]
This situation is widely expected when the detailed balance codition,
i.e. the time-reversal symmetry \cite{Onsager},
\[
P_0(\mib{x})W(\mib{x}\to\mib{x'})
=P_0(\mib{x'})W(\mib{x'}\to\mib{x}),
\]
or equivalently,
\begin{equation} \label{DB}
L(\mib{x},\mib{x'})P_0(\mib{x'})=L(\mib{x'},\mib{x})P_0(\mib{x}),
\end{equation}
is satisfied.
In this case the eigenvalues of $H$ are all real, and non-negative, if
the steady state is stable. Therefore, $\phi_0(\mib{x})$ is the
ground state.
\par
A useful example is the Fokker-Planck equation for a single, continuous
stochastic variable $x$,
\begin{equation} \label{FP}
 \DF{}{t}P(x,t)=-{\cal L}[x]P(x,t),~
{\cal L}[x]=-\DF{}{x}\left(F'(x)+\frac{\epsilon}{2}\DF{}{x}\right),
\end{equation}
which describes a one-dimensional Brownian motion in a potential
$F(x)$ with a small diffusion constant $\epsilon$.
By using its steady state solutions,
\[
 P_0(x)\propto \exp\left[ -2F(x)/\epsilon \right]~~\mbox{and}~~
\phi_0(x)\propto \exp\left[ -F(x)/\epsilon \right],
\]
we find the continuous variable version of the above formulations,
\begin{equation}
 {\cal H}[x]=\frac{1}{\epsilon}\left[
-\frac{\epsilon^2}{2}\DDF{}{x}+V(x)\right],~~
V(x)=\frac{1}{2}\left[ F'(x)^2-\epsilon F''(x)\right]. 
\end{equation}
Thus the Fokker-Planck equation is transformed into
a self-adjoint form of an imaginary-time Schr\"odinger equation,
\[
 -\epsilon\DF{}{t}\psi(x,t)=
\left[-\frac{\epsilon^2}{2}\DDF{}{x}+V(x)\right]\psi(x,t),
\]
and its eigenvalue problem results in a familiar one of the
quantum mechanics.
\par
Figure.1 shows an early application \cite{TIK}
to the so-called Kramers escape problem \cite{Kramers}.
The stochastic decay rate of the metastable state in a double-well
potential $F(x)$ is given by the first excited
eigenvalue $\lambda_1$ of the corresponding Schr\"odinger potential $V(x)$.
The first excited state is almost degenerated with the ground state for 
the small diffusion constant $\epsilon$.
\begin{figure}[t]
\begin{center}
\includegraphics[width=8cm]{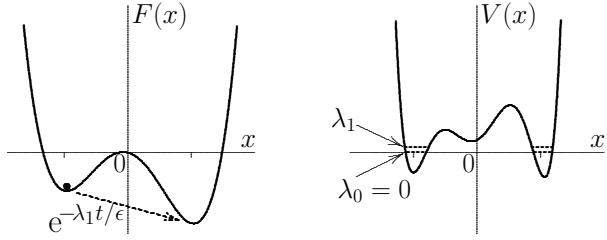}
\caption{\small Stochastic decay process of the metastable state.}
\vskip -0.5cm
\end{center}
\end{figure}
\section{Two-time conditional probability}
So far the quantum mechanical reformulation merely helps us to simplify
the eigenvalue problem of a given master equation.
None of the remarkable quantum mechanical phenomena appears, 
until we are concerned with the TTCP,
\begin{equation}
P(\mib{x},t|\xi,\ti;\xf,\tf),~~\ti\le t\le \tf~.~~~
\mbox{(~$;$ denoting `and')}
\end{equation}
By using the Markoffian property and the well-known equality of 
the simplest Bayes' theorem,
%
\[
P(A\cap B)=P(A|B)P(B)=P(B|A)P(A),
\]
%
repeatedly, the TTCP can be written in the following form with a pair
of the wave functions as
\begin{equation} \label{Twotime}
P(\mib{x},t|\xi,\ti;\xf,\tf)
=\frac{1}{\la\psif |\psii\ra}
\overline{\psi}(\mib{x},t|\xf,\tf)\psi(\mib{x},t|\xi,\ti),
\end{equation}
where the conjugate wave function $\overline{\psi}$~is related to the so-called ~\textit{posterior}~
conditional probability, ~$\overline{P}(\mib{x},t|\xf,\tf)$ for $t\le \tf$,~
by
\begin{equation}
\overline{\psi}(\mib{x},t|\xf,\tf)=\phi_0(\mib{x})^{-1}
\overline{P}(\mib{x},t|\xf,\tf),
\end{equation}
and obeys the \textit{backward} wave equation,
\begin{equation} \label{Backward}
\DF{}{t}\overline{\psi}(\mib{x},t) = \sum_{\mibsub{x'}}
H^{\dagger}(\mib{x},\mib{x'})\overline{\psi}(\mib{x'},t).
\end{equation}
Here $H^{\dagger}$ is the hermite conjugate of $H$, i.e. 
the transposed matrix in the present case. 
The eigensystem is common with the forward equation Eq.(\ref{Forward}),
when $H$ is hermite, i.e. real and symmetric as has been assumed here.\par
The denominator in Eq.(\ref{Twotime}) is the weight of overlap between the two wave functions defined by an inner product,
\begin{equation} \label{Inner}
\la\psif|\psii\ra
=\sum_{\mibsub{x'}}\overline{\psi}(\mib{x'},t|\xf,\tf)
\psi(\mib{x'},t|\xi,\ti).
\end{equation}
Of course this quantity is real, while the corresponding quantity
in the quantum mechanics is complex in general.
\par
Let us define the ket- and the bra-vectors by
\[
\ket{\psii(t)}=\{\psi(\mib{x},t|\xi,\ti)\}^{\dagger}~\mbox{and}~
\bra{\psif(t)}=\{\overline{\psi}(\mib{x},t|\xf,\tf)\}.
\]
Then the wave equations Eqs.(\ref{Forward}) and
(\ref{Backward}) are rewritten in the quantum mechanical form as
\begin{equation}
\DF{}{t}\ket{\psii(t)}=-H\ket{\psii(t)}~~\mbox{and}~~
\DF{}{t}\bra{\psif(t)}=\bra{\psif(t)}H.
\end{equation}
Henceforth, $H$ is called the Hamitonian.
\par
By using this pair of the Schr\"{o}dinger equations it is shown that the overlap integral,
$\la\psif|\psii\ra$ given by Eq.(\ref{Inner}) does
not depend on the current time $t$, i.e.
\[
\DF{}{t}\la\psif|\psii\ra=\la\psif(t)|H|\psii(t)\ra
-\la\psif(t)|H|\psii(t)\ra=0.
\]
Further it can be shown that this overlap integral has the following
properties in the two limits;
\begin{equation} \label{Limit0}
\left\{
\begin{array}{l}
(\mbox{i})~\displaystyle{\lim_{\tf-\ti\to\infty}}\la\psif|\psii\ra=1,\\
\\
(\mbox{ii})~\displaystyle{\lim_{\tf-\ti\to 0}}\la\psif|\psii\ra
=[\phi_0(\xf)\phi_0(\xi)]^{-1} \delta(\xf-\xi).
\end{array}\right.
\end{equation}
\par
Note that the two-time conditional expectation (TTCE) of a physical
quantity $Q$ with respect to TTCP defined by
\begin{eqnarray} \label{TTCE}
\la Q\ra_{\mbox{\scriptsize (i;f)}}
&=&\sum_{\mibsub{x'}} Q(\mib{x}')P(\mib{x}',t|\xi,\ti;\xf,\tf)\nonumber\\
&=&\frac{\la \psif(t)|Q|\psii(t)\ra}{\la \psif|\psii\ra},
\end{eqnarray}
has just the analogous form of the weak value in the quantum mechanics 
\cite{A-B}.\par
Thus the TTCP is a non-linear quantity composed 
of a product of a pair of the forward
and the backward wave functions, and cannot be described by a closed,
linear evolution equation.
Then it happens that the principle of the probability superposition
is violated and the interference of wave functions may occur.
However, its example is omitted here because none of 
nontrivial phenomenon from this view point has been found, yet.
The reason may be that the wave functions are always real and possitive
in the present case.
Let us discuss only the weak value in the rest.
\section{Stochastic model of classical Ising spins}
An example is a pair of the classical Ising spin $\sigma=\pm 1$
having an exchange interaction, 
\[
E(\mib{x})=-J\sigma_1\sigma_2,
\]
where $\mib{x}=(\sigma_1,\sigma_2)$.
Let us number the stochastic variable $\mib{x}$ in the order, 
$(1,1)$,~$(1,-1)$,~$(-1,1)$,~$(-1,-1)$ and choose the following 
transition matrices,
\begin{equation}
 W=\left(
\begin{array}{cccc}
0&1&1&0\\
p^2&0&0&p^2\\
p^2&0&0&p^2\\
0&1&1&0
\end{array}
\right)~~\mbox{or}~~
L=\left(
\begin{array}{cccc}
2p^2&-1&-1&0\\
-p^2&2&0&-p^2\\
-p^2&0&2&-p^2\\
0&-1&-1&2p^2
\end{array}
\right),
\end{equation}
where $p=\e^{-\beta J}$.
Evidently this transition matrix $W$ satisfies the detailed balance condition,
\[
\e^{-\beta E(\mib{x})}~W(\mib{x}\to\mib{x'})=\e^{-\beta E(\mib{x'})}~
W(\mib{x'}\to\mib{x}),
\]
at the steady state, i.e. the thermal equilibrium of a temperature parameter,
$\beta=1/k_{\mbox{\scriptsize B}}T$.
With use of the equilibrium distribution function,
\[
P_0(\mib{x})=\frac{1}{2(1+p^2)}~(1,~p^2,~p^2,~1)
~~\mbox{and}~~
\phi_0(\mib{x})=\frac{1}{\sqrt{2(1+p^2)}}~(1,~p,~p,~1),
\]
we find the corresponding hermite Hamiltonian,
\begin{eqnarray}
H&=&\left(
\begin{array}{cccc}
2p^2&-p&-p&0\\
-p&2&0&-p\\
-p&0&2&-p\\
0&-p&-p&2p^2
\end{array}
\right)
\nonumber\\
&=&(1+p^2)~\s0\otimes\s0
-(1-p^2)~\sz\otimes\sz-p~(\s0\otimes\sx+\sx\otimes\s0),
\end{eqnarray}
where $\sx$ and $\sz$ are the usual Pauli matrices and $\s0$ denotes the
two dimensional unit matrix $I_2$.
This is the Hamiltonian of a pair of \textit{quantum}
Ising spins with the exchange interaction in a transverse magnetic field.
\par
The eigenvalues and the eigenstates of this Hamiltonian $H$,
\[
\left\{
\begin{array}{l}
\lambda_0=0,~\lambda_1=2p^2,~\lambda_2=2,~\lambda_3=2(1+p^2),\\
\\
\ket{0}=\displaystyle{\frac{1}{\sqrt{2(1+p^2)}}}\left[~\ket{\!\ua\ua}~
+~p~\ket{\!\ua\da}+~p~\ket{\!\da\ua}~+~\ket{\!\da\da}~\right],
\\
\ket{1}=\displaystyle{\frac{1}{\sqrt{2}}}
\left[~\ket{\!\ua\ua}~-~\ket{\!\da\da}~
\right],\\
\ket{2}=\displaystyle{\frac{1}{\sqrt{2}}}
\left[~\ket{\!\ua\da}~-~\ket{\!\da\ua}~
\right],\\
\ket{3}=\displaystyle{\frac{1}{\sqrt{2(1+p^2)}}}
\left[~p~\ket{\!\ua\ua}~
-~\ket{\!\ua\da}-~\ket{\!\da\ua}~+~p~\ket{\!\da\da}~\right],
\end{array}
\right.
\]
can be easily obtained, where $\ket{0}=\ket{\phi_0}$, the ground state.
Here the familiar notations $\ua,\da$ are used for $\sigma=\pm 1$.
Note that the first excited state is almost degenerated with the ground
state for a small transition probability $p^2$.
\par
By using this eigensystem we can calculate the state vectors,
$\ket{\psii(t)}$
and $\bra{\psif(t)}$ for arbitrary initial and final states in just the same
manner of the elementary quantum mechanics except for the fact that
the time $t$ is imaginary.\par
An interesting example is the case where the initial and the final states
differ from each other, just like the case of the weak value.
For example, let
\[
\xi=(\ua\ua)~\mbox{at}~t=0~~\mbox{and}~~\xf=(\da\da)~\mbox{at}~t=\tf,
\]
that is,
\[
P(\mib{x},0)=(1,0,0,0)~~\mbox{and}~~\overline{P}(\mib{x},\tf)=(0,0,0,1),
\]
or equivalently,
\[
\ket{\psii(0)}=\sqrt{2(1+p^2)}~\ket{\!\ua\ua}~~\mbox{and}~~
\bra{\psif(\tf)}=\sqrt{2(1+p^2)}~\bra{\da\da\!}.
\]
By using the eigenvector expansion we obtain,
\begin{equation} \label{InitialWF}
\begin{array}{lcl}
\ket{\psii(t)}&=&\ket{0}+
\sqrt{1+p^2}~\e^{-\lambda_1 t}~\ket{1}+p~\e^{-\lambda_3 t}~\ket{3},\\
\\
\bra{\psif(t)}&=&\bra{0}
-\sqrt{1+p^2}~\e^{-\lambda_1(\tf-t)}\bra{1}+p~\e^{-\lambda_3(\tf-t)}\bra{3},
\end{array}
\end{equation}
and
\begin{equation} \label{Overlap}
\la\psif|\psii\ra=1-(1+p^2)~\e^{-\lambda_1\tf}
+p^2~\e^{-\lambda_3\tf}~(>0).
\end{equation}
\par
The TTCP is shown in Figure.2.
This result itself is very natural and well-expected,
all probabilities being always non-negative.
\begin{figure}[t]
\vskip 0.3cm
\begin{center}
\includegraphics[width=6cm]{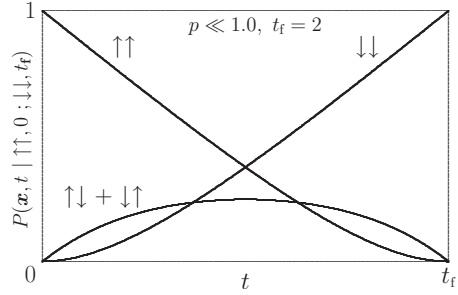}
\caption{\small Two-time conditional probability}
\end{center}
\end{figure}
\par
A strange behavior appears when we use the basis
$\{\ket{k},~k=0,1,2,3\}$, the eigenstates of the Hamiltonian
$H$ instead of the spin states $\{|\mib{x}\ra=|\sigma_1\sigma_2\ra\}$.
We can calculate the probability, i.e. the TTCE of the projection
operator $|k\ra\la k|$ onto each eigenstate $k$ in the same manner.
The result is given by
\begin{equation} \label{Projection}
\begin{array}{cclcl}
P(0,t)&=&\displaystyle{\frac{\la\psif(t)|0\ra\la 0|\psii(t)\ra}
{\la\psif|\psii\ra}}
&=&\displaystyle{\frac{1}{\la\psif|\psii\ra}}~,\\

P(1,t)&=&\displaystyle{\frac{\la\psif(t)|1\ra\la 1|\psii(t)\ra}
{\la\psif|\psii\ra}}
&=&-\displaystyle{\frac{(1+p^2)\e^{-\lambda_1\tf}}{\la\psif|\psii\ra}}
~~(<0~)~,\\

P(2,t)&=&\displaystyle{\frac{\la\psif(t)|2\ra\la 2|\psii(t)\ra}
{\la\psif|\psii\ra}}
&=&0~,\\
P(3,t)&=&\displaystyle{\frac{\la\psif(t)|3\ra\la 3|\psii(t)\ra}
{\la\psif|\psii\ra}}
&=&\displaystyle{\frac{p^2\e^{-\lambda_3\tf}}{\la\psif|\psii\ra}}~.
\end{array}
\end{equation}
The fictitious negative probability is found in $P(1,t)$.
Of course the completeness of the probability,
\[
\sum_{k=0}^3 P(k,t)=1,
\]
is satisfied evidently because of Eq.(\ref{Overlap}).
\par
A related unusual behavior to this fact is the abnormal enhancement
of some observables.
For example, if we calculate the TTCE of a quantity,
\begin{equation} \label{Nondiagonal}
M_x=\frac{1}{2}(\sx\otimes\s0+\s0\otimes\sx),
\end{equation}
an abnormal behavior
\[
\la M_x\ra_{\mbox{\scriptsize (i;f)}}
=\frac{1}{\la\psif|\psii\ra}\left[\displaystyle{\frac{2p}{1+p^2}\left(1-p^2\e^{-\lambda_3\tf}\right)
-\frac{1-p^2}{1+p^2}\left(\e^{-\lambda_3 t}+\e^{-\lambda_3(\tf-t)}\right)}\right]
>1,
\]
is found for sufficiently small $p$ and $\tf$.
An example is shown in Figure.3.
\begin{figure}[t]
\vskip 0.5cm
\begin{center}
\includegraphics[width=6cm]{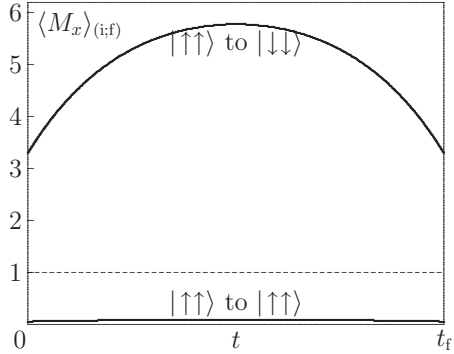}
\caption{\small Abnormal and normal TTCE of $M_x$ for
$p=0.2$ and $p^2\tf=0.01$.}
\vskip -0.5cm
\end{center}
\end{figure}
Note that the natural norm of $M_x$ must be less than 1,
because the eigenvalues of $M_x$ are $\{-1, 0, 0, 1\}$.
When the transition rate is very small, 
i.e. $p^2 \tf\ll 1$, we find
\[
\la M_x\ra_{\mbox{\scriptsize (i;f)}}\gg 1.
\]
A plain reason of this singular behavior is that the overlap integral
$\la\psif|\psii\ra$ in the denominator may be expected to be very small
owing to (ii) of Eq.(\ref{Limit0}), whenever
the initial and the final states differ from each other, i.e. $\xi\ne\xf$.
This means that to reach $\xf=(\da\da)$ starting from $\xi=(\ua\ua)$
in a given time occurs scarcely and is far from the main flow
of the conditional probability.
On the contrary none of such strange behaviors are found
when $\xi=\xf$, e.g. $\xi=\xf=(\ua\ua)$.
The result for the latter case for the same parameters as the upper
abnormal case is shown by the lower curve in Figure.3,
its maximum being $\sim 0.09$ at $t=\tf/2$ and minimum $\sim 0.05$ at $t=0$
and $\tf$.
\par
In Figure.4 the TTCE of another
quantity $A=\sx\otimes\sx$ are shown also.
Note that $A$ is commutative with $H$ and a conserved quantity.
Then the horizontal axis in this figure shows a parameter 
of the transition probability, not the time.
\begin{figure}[bt]
\vskip 0.5cm
\begin{center}
\includegraphics[width=6cm]{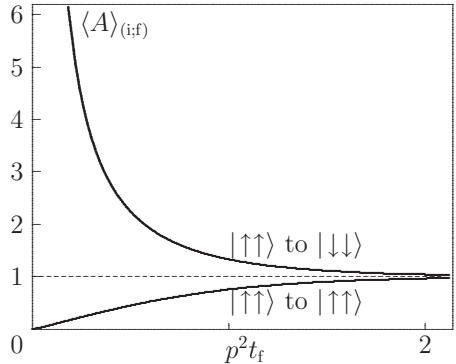}
\caption{\small Abnormal and normal TTCE of $A=\sigma_x\otimes
\sigma_x$ for $p=0.2$.}
\vskip -0.5cm
\end{center}
\end{figure}
\section{Extension of TTCP to a density matrix}
It should be noted that the physical quantities $M_x$ and $A$ are
 \textit{non-diagonal} in the spin-state representation and have no 
corresponding quantities in the classical Ising spin system.
They are related to the transition rate of the stochastic Ising spin.
In order to calculate the expectations of such non-diagonal quantities
we need an extension of the TTCP to the two-time conditional (TTC)
density matrix defined by
\begin{eqnarray} \label{Density}
\rho_{\mbox{\scriptsize (i;f)}}(t)
&=&\frac{1}{\la\psif|\psii\ra}~
|\psii(t)\ra\la\psif(t)|\nonumber\\
&=&\frac{1}{\la\psif|\psii\ra}
\sum_{\mibsub{x},\mibsub{x'}}
\overline{\psi}(\mib{x'},t|\xf,\tf)\psi(\mib{x},t|\xi,0)~
\ket{\mib{x}}\bra{\mib{x'}}.
\end{eqnarray}
From the definition Eq.(\ref{Inner}) of the overlap integral
$\la\psif|\psii\ra$, it is evident that
\[
\mbox{Tr}~\rho_{\mbox{\scriptsize (i;f)}}(t)=
\frac{1}{\la\psif|\psii\ra}
\sum_{\mibsub{x}}
\overline{\psi}(\mib{x},t|\xf,\tf)\psi(\mib{x},t|\xi,0)
=1.
\]
It should be noted, however, that diagonal elements of this
density matrix are not always positive as is shown by Eq.(\ref{Projection})
in Sec.4, when it is diagonalized by using the basis
$\{|k\ra,k=0,1,2,3\}$, the eigenstates of the Hamiltonian $H$.
\par
With use of this density matrix the definition Eq.(\ref{TTCE})
of the TTCE is extended as
\[
\la Q\ra_{\mbox{\scriptsize (i;f)}}=
\mbox{Tr}~\rho_{\mbox{\scriptsize (i;f)}}Q.
\]
This definition of the TTCE results in the classical one,
if $Q$ is diagonal.\par
The notion of this density matrix has not been used in the conventional
classical stochastic process.
It should be emphasized, however, that this quantity is within
a scope of the classical stochastic process itself,
because the wave functions,
$\psi$ and $\overline{\psi}$ in Eq.(\ref{Density}) are related to the
forward and the posterior, classical conditional probabilities,
respectively.
In addition, we have an alternative expression for $\overline{\psi}$,
\begin{equation} \label{Inverse}
\overline{\psi}(\mib{x'},t|\xf,\tf)
=\psi(\mib{x'},\tf|\xf,t)~
\left(=\phi_0(\mib{x'})^{-1}P(\mib{x'},\tf-t|\xf,0)\right),
\end{equation}
or equivalently,
\begin{eqnarray}
\overline{P}(\mib{x'},t|\xf,\tf)P_0(\xf)&=&P(\xf,\tf|\mib{x'},t)P_0(\mib{x'})
\nonumber\\
&=&P(\mib{x'},\tf|\xf,t)P_0(\xf),
\end{eqnarray}
for $t\le\tf$ due to the time-reversal symmetry corresponding to the 
detailed balance.
Then the density matrix Eq.(\ref{Density}) can be written as
\begin{equation}
\rho_{\mbox{\scriptsize (i;f)}}(t)
=\frac{1}{\la\psif|\psii\ra}
\sum_{\mibsub{x},\mibsub{x'}}
\displaystyle{
\frac
{P(\mib{x'},\tf -t|\xf,0)P(\mib{x},t|\xi,0)}
{\phi_0(\mib{x'})\phi_0(\mib{x})}
}
\ket{\mib{x}}\bra{\mib{x'}}.
\end{equation}
This fact means that we can define the TTC density matrix with only a pair of
the usual, forward conditional probabilities for two individual 
initial states, $\xi$ and $\xf$.
\section{Summary and discussions}
Except for the facts that the time is imaginary and the wave function is 
always real and positive, the classical stochastic process can be 
described in an analogous form of the quantum mechanics, if we use 
the TTCP.
For example, the abnormal behaviors of the weak value in the quantum
mechanics are emulated.
Note that the TTCP and the weak value are always
real in the present classical case.
Therefore, the origin of such abnormal behaviors is clearer than the 
quantum mechanical case where complex quantities appear.
In addition, if we have not the explicit solution of the eigenvalue problem,
we may calculate the weak value at least with use of the Monte-Carlo
simulation which is often used to investigate the stochastic model.
\par
The importance of the weak value in the quantum mechanics
is that it is related to the new notion of the
weak measurement without disturbing the quantumu state.
An analogous notion of the latter in the classical stochastic process 
has not been found yet.
\par
\bigskip
\noindent
{\bf Acknowledgements}
\par
\bigskip
This work is supported by Open Research Center Project
for Private Univercities: Matching fund subsidy from
MEXT of Japan.


\end{document}